\documentclass{elsarticle}

\usepackage{lineno,hyperref}
\modulolinenumbers[5]

\journal{International Journal of Human-Computer Studies}

\biboptions{authoryear}

\usepackage{booktabs}
\usepackage[table,xcdraw]{xcolor}
\usepackage[final]{changes}

\begin{document}

\begin{frontmatter}

\title{The Users' Perspective on the Privacy-Utility Trade-offs in Health Recommender Systems\tnoteref{mytitlenote}}
\tnotetext[mytitlenote]{Full article available on \href{https://doi.org/10.1016/j.ijhcs.2018.04.003}{https://doi.org/10.1016/j.ijhcs.2018.04.003}. \\ \copyright 2018. This manuscript version is made available under the CC-BY-NC-ND 4.0 license http://creativecommons.org/licenses/by-nc-nd/4.0/}

\author{Andr\'{e} Calero Valdez  \&  Martina Ziefle}
\address{Chair of Communication Science, Human-Computer Interaction Center}

\author[mymainaddress]{RWTH Aachen University, Germany}

\cortext[mycorrespondingauthor]{Corresponding author}
\ead{calero-valdez@comm.rwth-aachen.de}

\address[mymainaddress]{Campus Boulevard 57, 52074 Aachen, Germany}

\begin{abstract}
Privacy is a major good for users of personalized services such as recommender systems. When applied to the field of health informatics, privacy concerns of users may be amplified, but the possible utility of such services is also high. Despite availability of technologies such as k-anonymity, differential privacy, privacy-aware recommendation, and personalized privacy trade-offs, little research has been conducted on the users' willingness to share health data for  usage in such systems. In two conjoint-decision studies (sample size $n=521$), we investigate importance and utility of privacy-preserving techniques related to sharing of personal health data for k-anonymity and differential privacy. Users were asked to pick a preferred sharing scenario depending on the recipient of the data, the benefit of sharing data, the type of data, and the parameterized privacy. Users disagreed with sharing data for commercial purposes regarding mental illnesses and with high de-anonymization risks but showed little concern when data is used for scientific purposes and is related to physical illnesses. Suggestions for health recommender system development are derived from the findings.
\end{abstract}


\end{frontmatter}


\section{Introduction}
Modern societies are burdened with demographic change. Low birthrates, high life-expectancy, and urbanization affect the availability of workforce and health care---and not \replaced{in a positive way}{positively so}. In rural areas, medical care becomes increasingly unavailable, due to doctors seeking jobs in cities. Here in particular, population age shifts to the very old, as younger people move to urban areas to find employment, education, and opportunities~\citep{wilson2009inequitable}. \added{Can the utilization of health recommender systems help alleviate these challenges?}

By simplifying access to digital medical services, public health could benefit~\citep{brodie2000health}.
Digitizing health data and utilizing computational power could provide a relatively easy access to personalized medicine. 
This would also improve public health surveillance---the ongoing, systematic collection, analysis, interpretation, and dissemination of data regarding health---and improve policy making. 
One approach to personalized health lies in the use of recommender systems, or, more specifically, health recommender systems.


But the question is, how can recommender systems be applied to the sensitive field of health? How can ``finding interesting'' items be relevant for health care? The algorithms used in recommender systems use similarity of users to identify ``matching'' items in relation to the similarity criterion. This criterion can be exchanged for any health-related criterion. For example, by using medication data and further patient data, recommender systems could be used to suggest medication that has less side effects \citep{zhang2016predicting}. Health recommender systems could suggest therapies that better match patients' dispositions and adherence behaviors \citep{hidalgo2014glucmodel,esteban2014tplufib}. By suggesting items that have been satisfactory for patients with a similar health status or disease history, \deleted{a} first access to personalized medicine could be achieved.

This benefit does however come with a \textit{privacy trade-off}.
Recommender system approaches require knowledge about the users. This knowledge is often created implicitly (by buying or reading) or explicitly (by asking the user). However, in an e-commerce setting this type of information could be described as ``low-risk'' information. The information that a person likes ``A Hitchhiker's Guide to the Galaxy,'' might not be a strong invasion of privacy (as it represents quite generic information about the person). 
The perceived costs \added{or risks} of sharing information (``telling the world I like A'') is smaller than the expected utility (``discovering good product B''). 
Regardless of the quality of the underlying recommendation algorithm, users can judge whether this privacy trade-off\deleted{s} is desirable.
The aforementioned cost is to a large extent the risk that one links to the fact that the world knows that one likes a particular \replaced{choice}{product}. 
The idea of \textit{privacy-aware algorithms} \citep{alvim2011differential} refers to algorithms that allow to ensure privacy with respect to a given threshold. It allows to parameterize privacy. 
Depending \replaced{on}{of} the required information accuracy, they may decide to use lower quality of data, thus anonymizing data, while allowing algorithms to still work. But how to choose the ``right'' level of privacy, especially in the health domain, which deals with highly sensitive and personal or even intimate information, on the one hand, and requires highly accurate information on the other? Should we rely on technical considerations like specific algorithms, as proposed by~\cite{lee2011much}, or should we ask the users? And if so, how do we ask?

The potential of combining Information and Communication Technologies (ICTs) such as recommender systems and health is obvious, not only for patients but also for the care personnel and the health care system. But can we reach a high public acceptance for health recommender systems without understanding the position of patients, their perceived barriers and benefits of sharing information? In how far does the type of medical data (mental vs. physical illnesses) or the type of benefit users receive modulate patients' decisions to share information? 

\paragraph{Main Contribution}
In this paper, we investigate how privacy is perceived from users' perspective when health data is stored for different uses and different benefits in a recommender system. Note that we did not use a specific implementation, but we concentrated on users' perceptions and simulated their decision to share medical data.
We follow our framework proposed in \cite{CaleroValdez:2016:HRS:2959100.2959158} that suggests using a holistic perspective for research in health recommendation.  
\deleted{For this purpose} We look into how users perceive the privacy utility trade-off in different usage contexts.
We measure how much different levels of privacy are worth in different usage scenarios. For this purpose, we compare the use of k-anonymity with the use of differential privacy in two quantitative user studies investigating the user's attitudes towards privacy in health recommender systems. 
\section{Related Work: Why We Need Research on Privacy in Health Recommender Systems (HRS)}
\begin{figure}[htb]
\centering
\includegraphics[width=0.6\textwidth]{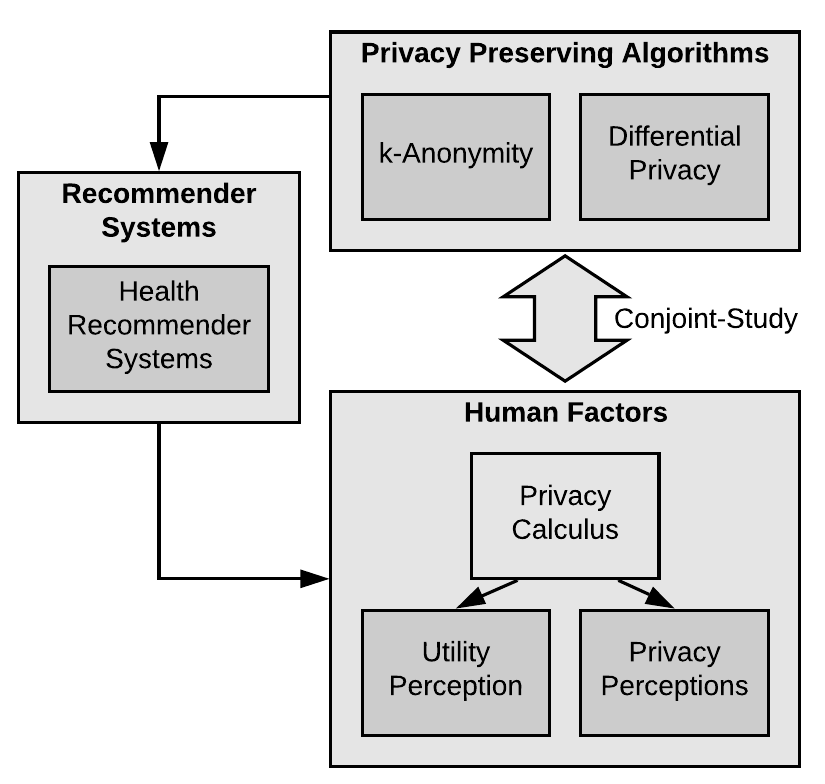}
\caption{Overview of the related fields of work and where our conjoint studies aim at.}
\label{fig:sota}
\end{figure}

The benefits of health informatics and health recommender systems are undeniable---for the users, professionals, and societies as a whole (see Section~\ref{benefits}). As with any complex system, understanding its parts and their interconnection is critical.
To understand the interplay of users' attitudes, health recommender systems, and privacy, we first investigate the need to consider the users' perspective by looking at the influence of users' attitudes on acceptance (see Section~\ref{acceptance} and Fig.~\ref{fig:sota}). Without acceptance and the willingness of users to share their information and allowing recommender system to use it, even perfect algorithms are meaningless. 
As one core issue for users is data privacy, we then look at how privacy is relevant to both recommendation algorithms and users' attitudes. Information privacy plays a \replaced{distinctive}{peculiar} role in the evaluation of the users' privacy calculus regarding personal health records (see Section~\ref{privcalc}). Therefore, some  systems have integrated  so-called privacy-aware algorithms or privacy preserving algorithms (see Section~\ref{diffpriv}), even \deleted{though mostly} in non-health-related contexts. These concepts have been integrated into recommender systems (see Section~\ref{privacyRS}), however the complex field of health recommender systems is very diverse and has only seen few privacy-aware implementations \added{so far} (see Section~\ref{HRS}). 

\subsection{Benefits of Health Informatics}
\label{benefits}
\deleted{Without any doubt, h}Health informatics represents a huge technical and social benefit for countries, societies, and individuals \citep{pagliari2007potential,martin2014big}. Health informatics profits from the enormous advances in \replaced{ICT}{information and communication technologies}\deleted{across the Internet} and the digitalization of health data. It allows a fast, seamless, and continuous collection, analysis, interpretation as well as dissemination of health data \citep{german2001updated}. The benefits of using electronic health data relate to a multitude of advantages in terms of time-critical and accurate diagnosis and treatment, e.g., the identification of emerging diseases, the identification of populations at risk, health behaviors that are critical, as well as detection of epidemics \citep{detmer2003building}. However, not only care givers or health specialists profit from the availability of public health data but also the patients themselves \citep{ferguson1997health}. Digital health services can inform patients very accurately about the status of their disease, may aid patients through self-care, provide fast access to medical care also from remote places, connect them with other patients and care personnel, and allow shared decision making. \deleted{Not only is patients' safety increased by having access to health records and digital health services.} In addition, patients' health awareness and health motivation could be heightened, as patients bear responsibility for their own health by being an integral part of digitally assisted health care~\citep{holzinger2010chances}.

Although the number and quality of studies remains limited, existing research suggests improvements in communication and trust between patients and professionals, confidence in self-care, compliance in chronic disease \added{management}, and accuracy of records~\citep{li2010securing,lymberis2003intelligent,kalra2006confidentiality}. Patients particularly value online reordering of prescriptions, laboratory results, disease management plans, trend charts, drug lists, and secure messaging~\citep{ferguson1997health}. 
Empirical evidence indicates that most patients would like to be able to access their \added{own} personal health records~\citep{ferguson1997health,kovats2005governance,kowalewski2015like}. \deleted{The emergence of mobile and wireless applications that allow remote submission of data to a shared record storage, offer new possibilities for patient monitoring and real time decision support. Additionally, electronic records may help to promote partnership between carers and health professionals through sharing information, or allow relatives to monitor the care and recover of elderly parents or children in hospitals from a distance.} 
Still, however, there are significant \replaced{obstacles concerning}{drawbacks in the}  confidentiality, the security, and the privacy of such sensible health data.

\subsection{Acceptance of Medical Technology and Digital Services}
\label{acceptance}
\deleted{Despite the possible benefits provided by health informatics for health-care and medical safety, most of the systems impose one problem: they ignore privacy of the persons providing their data. Privacy is the right of protection of people's personal information, i.e., confidentiality, anonymity, self-determination, freedom of expression, and personal control of data. }


Privacy is one important determinant of any health informatics system acceptance~(e.g., \citealt{kowalewski2015like}). The understanding of these determinants that affect technology acceptance is essential for its successful adoption \citep{mandl2001public}. The most influential and best-established theoretical approach to explain and predict the adoption of technologies is the Technology Acceptance Model (TAM, cf. \citealt{davis1986technology}), which was also adapted to the health care context \citep{holden2010technology}. However, these models cannot be easily transferred and applied to the design and implementation of health recommender systems for several reasons. 
First, technology acceptance models focus on the evaluation of complete technical systems or applications. They do not provide information about the evaluation of single technical characteristics of a system. Accordingly, practical design guidelines for health recommender systems design cannot easily be derived, e.g.~``what data should be stored'', ``which algorithm should be used'', or ``for what purposes should data be further used?''. 
Second, in \replaced{in practice}{user} studies about health informatics system acceptance, the design process of the product is often finished and users are being confronted with technically mature prototypes~\cite{pagliari2007design}, where only marginal changes can be made. In order to optimally support the acceptance of health recommender systems, users' needs and 
requirements should be assessed as early as possible in the systems' design life cycle to steer acceptance and usage conditions in line with the technological development. 
Third, technology acceptance and the willingness to use technical systems in sensible domains is a multi-dimensional concept that is highly context-specific. Depending on the usage context, identical technical systems or functions are perceived differently by users. 
For example, an evaluation of recommender system for e-commerce~\citep{pu2012evaluating} focuses other aspects relevant for acceptance. The risks associated with the misinformation of a ``bad recommendation'' might influence adoption differently than the risk of a bad shopping recommendation. Little work on evaluating health recommender systems has been done, as the whole research field is quite young.

\subsection{Privacy Calculus}
\label{privcalc}
\added{Privacy is the right of protection of people's personal information} \citep{kovats2005governance},\added{ i.e., confidentiality, anonymity, self-determination, freedom of expression, and personal control of data.}
The term privacy has several definitions, differing between fields of research. In computer science, privacy is strongly connected to an adversarial model and how much data about individuals can be obtained from either single or repeated data requests. 
In the social sciences, however, some definitions \replaced{regard privacy as}{relate them to} rights, states ,or in other definitions even as a commodity~\citep{zeissig2017online}.
Privacy is referred to as the ``selective control of access to the self''~\citep{altman1976conceptual} or, with respect to information privacy, ``the ability to control who gathers and disseminates information about one's self or group and under what circumstances'' \citep{burgoon1989maintaining}. This definition includes the main aspect of the privacy calculus: selection. The user controls privacy by deciding which information about himself to retain and which to release. Privacy is not the protection of all the users' information but the selective purpose- and usage-bound release of information to a defined group of recipients. The users evaluate utility of information release against possible privacy risks~\citep{vervier2017perceptions}.

Empirical studies regarding privacy from a user's perspective have mostly investigated \textit{privacy concerns} 
\citep{belanger2011privacy,li2011empirical,smith2011information}.
Privacy concerns refer to the individuals experience of dissonance between one's privacy expectations or desires and the actual (technical) privacy. These concerns may be very specific (e.g., privacy concerns \replaced{when using}{in the use of} facebook) or generic (e.g., privacy concerns \replaced{when using}{in the use of} the Internet). Privacy concerns are often measured as a multi-dimensional construct. \cite{malhotra2004internet} divide privacy concerns into control (or lack thereof), awareness (of privacy-related techniques), and collection (techniques of data). Other authors have used sub-dimensions, such as unauthorized secondary use, error (and leakage of data), and improper access (by hackers or criminals).

\subsection{Privacy Preserving Technologies}
\label{diffpriv}
The term ``privacy preserving technology'' refers to a set of methods \replaced{to ensure}{of ensuring} that privacy concerns are respected in databases. The aim is to maximize utility of a database while minimizing the risk to identify individual records. 
The simplest approach of anonymization is to remove data from individual releases of data. This should ensure that every set of columns (or features) occurs at least k-times, leading to the concept of \textbf{k-anonymity}~\citep{bayardo2005data}. This means that at least k users exist for whom the released data are completely equal. Thus, any individual is anonymous in a set of k individuals. 

Yet, depending on the diversity of data, \textit{k-anonymization} may still lead to de-anonymization of individual users in homogeneous data sets or with background knowledge of data \citep{machanavajjhala2007diversity}. This makes it essential to ensure that the database contains a sufficiently diverse set of data to protect users. The \textit{l-diversity} approach works by adding intra-group diversity. This helps reducing attacks based on homogeneity within groups. But even l-diversity approaches can be attacked if distribution of heterogeneity in data is not respected adequately \citep{li2007t}. Applying the concept of t\textit{-closeness} improves on this pitfall. Still, the problem of combining different data sets from sequential or independent releases persists.

The concept of \textbf{differential privacy} refers not to how data is stored in the database, but to how data is perturbed in a database request, depending on the the risk of exposing individuals in a request~\citep{dwork2006calibrating}. Differential privacy ensures that the utility of queries is maximized regarding the statistical properties of the dataset while minimizing the risk of exposing an individual. Algorithms that follow the concept of differential privacy have privacy parameter $\epsilon$ that determines the trade-off \replaced{between}{of} privacy and utility for a request. This yields a fixed privacy budget in cumulative requests. Every further request increases the amount of ``leakage'' and thus needs more perturbation or more datasets to ensure the same low privacy parameter. 
In a fixed setting, the risk of identification for any given individual thus depends on individual factors, the exceptionality of the individual's data, and the sample size. Differential privacy minimizes the individual's exposure according to these criteria.    

A broad set of privacy conserving algorithms have been reviewed by \cite{aggarwal2008general}. The authors point out that by reducing data fidelity, losses in utility are often inevitable. However, this trade-off of utility and anonymity can be maximized \replaced{with}{to} different optimization criteria~\citep{li2009tradeoff}. 
Personal privacy needs might differ between users and thus approaches of personalized privacy preservation could help maximize individual utility~\citep{chellappa2005personalization}. In this social context, trade-off refers to the individual perception and weighing of criteria that justify decision making. Thus, we focus on the ``perceived privacy-utility trade-off", e.g., users might decide to take the risk of data sharing as they perceive to \added{be} in control. Likewise, users might decide to share data because the temporary benefit is higher for them than the potential risk.
This utility might differ between individuals and correlate with their willingness to share. In health scenarios, individual sharing is nevertheless peculiar. The release of health data might also affect other individuals \citep{dankar2012application} and de-anonymize them as well. For example, giving away genomic-data also affects the privacy of other family members. 

\subsection{Privacy in Recommender Systems}
\label{privacyRS}
Some approaches to integrate privacy-aware algorithms into recommender systems exist 
\citep{mcsherry2009differentially}, showing very good performance in sufficiently large datasets. One may argue that using a movie recommender is relatively harmless, but personalized movie preference data comes with a risk of de-anonymiziation. \cite{ramakrishnan2001privacy} were able to reveal even political stances for individual users from seemingly innocent data sets.

The problem lies in the sparsity of data in recommender systems. \cite{narayanan2008robust} have shown a robust de-anonymization approach for sparse data sets. Methods like k-anonymization can hardly be applied to very sparse data sets. The aim of the underlying algorithms is to maximize personal utility of recommendation, whatever criterion utility may be. This contradicts the challenge of anonymization where individual preference hides among k other individuals.

Modern approaches integrate concepts from differential privacy and randomized perturbation into recommender systems \citep{liu2017differential}, guaranteeing privacy while maintaining high accuracy. Cryptographic approaches for social recommendations have been tested successfully by \cite{liu2015secure}, shielding recommendation data from social network data between two unrelated sources.

The use of recommender systems in the field of health requires us to take the bigger picture into account \citep{valdez2016state}. Many patients suffer from ``rare diseases,''\footnote{As there are a lot of rare diseases, it can still be possible for a large proportion of patients to suffer from different rare diseases.} so sample sizes for the individual diseases are small, with high de-anonymization risks.
The promised benefit of improved health and health care might cause users to see utility in data-disclosure, but, at the same time, they might overlook or underestimate possible de-anonymization risks. 
Explaining these risks to users might reduce their trust in the system, even if the system honestly conveys the risks of entering personal information \citep{knijnenburg2013making}. 
New forms of communicating risks and utility of privacy preserving techniques also need to be developed. Further, personal privacy needs differ between users and thus approaches of personalized privacy preservation help maximize individual utility \citep{chellappa2005personalization}, also for health recommender systems.

\subsection{Health Recommender Systems (HRS)}
\label{HRS}
\deleted{The field of health recommender research is quite young.  As of May 7th 2017 only 40 \it{(53 article match the search term but 13 are not related directly to the topic)} articles are found when searching for the terms ``recommender system'' and `` health'' in web of science. The oldest article is from 2007 and the most cited article has only 11 citations. In a short review on health recommender systems by } 
\deleted{, the increasing importance of Health Recommender Systems (HRS) is stated. HRS are tools used to aid decision making processes in all health care services. They show a potential to improve the usability of health care devices by reducing information overload from medical devices and software and thus improve their acceptance.}

Recommender systems are used for different purposes in health informatics. There are typically two target users for a HRS.
 \cite{wiesner2014health} discern between systems for  health professionals as end-users and systems for patients as end-users.
For health professionals, recommender systems are typically used to improve information access either for a specific case, clinical guidelines, or research articles. For patients, recommender systems should either provide high quality health information in an intelligible fashion or alternative procedures for illnesses, fitness, or nutrition. 
The aims of HRS include providing relevant information to end-users that is trustworthy\deleted{, as in the work of}~\citep{wiesner2014health}, lifestyle change recommendations that are actionable~\citep{farrell2012intrapersonal}, and improving patient safety \citep{roitman2010increasing}. \deleted{Improving patients safety could, for example, be achieved by providing information on possible side effects or interactions of drugs to reduce risks.}

Besides this \textit{recipient}-focused categorization, the recommended \textit{item} can also differ. Looking at HRS from this perspective yields research focused on recommending relevant information (the classic recommender scenario), diagnostics, therapies, and fitness or health behavior.

\paragraph{Information Access}
\replaced{Little of the available health information is actually utilized by patients}{In their research statement Fernandez et al.~state, that of all the available information on health topics, little is used from a patients' perspective, as access to this information is not properly enabled.}~\citep{fernandez2009challenges}.
\cite{turoff2008future} \replaced{apply}{addressed this problem by applying} a social recommender system that uses a collaborative filtering approach to find \textit{gray literature}. \deleted{This setup is used by professionals mostly to improve information access.}
\replaced{Similarly, }{The use of social networks as a means of providing target-specific information has been investigated by} \cite{song2012anonymous} \added{tried }to identify \textit{health social networks} relevant for a patient. \deleted{The authors also aim to preserve privacy by using algorithms based on discriminant words for finding options.} No direct patient data is used in the\added{ir} recommender system. 
\deleted{Song et al. 
recommend a social network of related parents for information exchange and also put efforts into preventing leakage of sensitive information.} 
\added{Similar approaches were applied to Youtube videos, expert finding, and the utilization of personal health records}~\citep{rivero2013health,kerschberg2014role,ati2015integration}.
\deleted{Rivero-Rodriguez et al. 
use a recommender system to enrich YouTube videos with medical information. The purpose of this approach is to make \textit{trustworthy information} more accessible for patients.
Kerschberg 
has implemented a method to utilize personal health records (PHR) to improve \textit{search} in health related areas. In contrast Guo et al. 
help identifying medical experts in a field by analyzing scientific literature. \textit{Key opinion leaders} regarding a disease are identified as authors that publish on the subject---an approach also implemented by Narducci et al.
. 
The use of personal health records has also been established as an integration for knowledge bases, which are then used in recommender systems Ati et al. 
.}

\paragraph{Diagnostics}
Besides improving information access, recommender systems are also used to help with diagnosis. \cite{thong2015hifcf} help professionals by providing fuzzy picture clustering and recommendation for possible illnesses, thus improving diagnostic accuracy. \deleted{A similar approach is done by Hussein et al. 
but for chronic diseases.} \cite{pattaraintakorn2007web} have used rough sets, survival analysis, and patient data to recommend clinical \textit{examinations} to improve early diagnostics. 
\cite{lafta2015intelligent} have used techniques from recommender systems to predict short-term \textit{risk} for heart disease patients from personal health records. The field of diagnostics is naturally high in risk; thus most approaches are more decision-support tools than recommendation tools. 

\paragraph{Therapies}
When a diagnosis is \replaced{made}{reached}, the selection of therapy comes next. Most systems focus on the reduction of side-effects and improvement of quality of life for evaluation. Recommender systems at this stage of care typically include information of previous stages. 
\deleted{Liu et al. 
improve information access regarding the identification of clinical care pathways. Their recommender systems is used to suggest causal links between user-diversity factors, co-morbidities and possible \textit{interventions}.}

Many approaches have been used to predict and thus \textit{prevent side-effects} and interactions of medication \citep{pinto2015predicting}. 
These methods are particularly successful when patient data is used \citep{grasser2016application}. \cite{zhang2015framework} predict side-effects by using a hybrid recommender systems based on personal health records and the experience of patients. 

\deleted{
Different algorithms are applied in HRS. Zhang et al. 
improved their system by using machine learning---neural networks or in particular Boltzman machines. Bao et al. 
have successfully applied support vector machines. 
}

\cite{chen2015orderrex} analyzed patient data of 18,000 patients. Their top-10 recommendations could be improved by including personal health records. However, \deleted{they also argue that} higher precision not necessarily means more correct decisions. By using past-data as evaluation, it could be that common therapies are preferred over ``better'' therapies. A similar concern has been voiced by \cite{hao2016comparative} who conducted \textit{risk prediction} using collaborative filtering. They argue that classification approaches could be better suited if \replaced{they}{it} existed \added{for the individual disease}.  
Collaborative filtering approaches may be more successful in predicting the individual utility of a therapy than the healthiest option \citep{parimbelli2015collaborative}. 

\deleted{A unique approach has been applied by Hamed et al.
, who use a Twitter-based recommender system. Twitter data on illnesses is scraped and by using a doctor-in-the-loop approach, \textit{alternative treatments} for illnesses are suggested. The  questionnaire based recommendation has a ``novelty'' focus on finding new treatments unknown to the users.} 

\deleted{Beyond these singular therapy recommendations, Hidalgo et al. 
implemented a recommender systems for the control of diabetic health and combined it with an e-learning system, case-based reasoning, patient data and doctor preferences. 
A system that provides peer support for diabetics was developed by 
Chomutare et al. 
}

\paragraph{Health Behavior}
Besides treating illness conditions, a large field for health recommendation is health behavior recommendation. 
\cite{sasaki2013walking} use recommendation to suggest \textit{walking routes}. These routes are selected for safety, amenity, and walkability. Users who need to avoid steep slopes or require resting locations can find matching routes by supplying their user data. Similar approaches have been done for running routes  
\citep{ISI:000393155700013} and general mobile activities
\citep{torres2015mobile}.

\deleted{Esteban et al. 
have developed a recommender system to help with lower back pain. The system suggests individualized exercises according to the users' previous behavior, needs, and requirements.}

Other lifestyle change recommendations focus on suggesting users how to improve their eating \citep{rokicki2015s,elsweilerbringing,espin2015nutrition,trattner2017estimating,harvey2015automated,said2014you}, exercising, sleeping behavior, or support them in quitting smoking~\citep{sadasivam2016impact}. 

\paragraph{Privacy in HRS}
The concern with many of these systems is that personal health records or other user data that is particularly sensitive is leveraged for recommendation and thus possibly exploited for unauthorized secondary use (e.g., adjusting insurance tariffs). Further, when using a recommender system, a performance metric must be defined. Ulterior motives of the developers could exist (e.g., selling a particular drug or upselling more expensive therapies) and be included in recommendation metrics. \deleted{And} By understanding the needs of patients, those needs can also be exploited.

\cite{hu2016personal} address this by anonymizing personalized information from personal health records before using them in a recommender system. \cite{kandappu2014privacycanary} go a step further and allow users to set up their own privacy-utility preference \added{which is} applied in their recommender system. However, this system is not a health recommender system. And it remains questionable how users are able to determine a good trade-off for health-related questions.
In particular, as it is still unknown how good recommendations in a health scenario can become \citep{said2012users}.
Overall, health recommender systems require information about the patient, user, and the applied context to operate successfully. So far, no research has investigated the users' preference on what they are willing to share, for which purposes, and with whom.

\section{Empirical Methodological Approach}
\begin{figure}[htb]
\includegraphics[width=\textwidth]{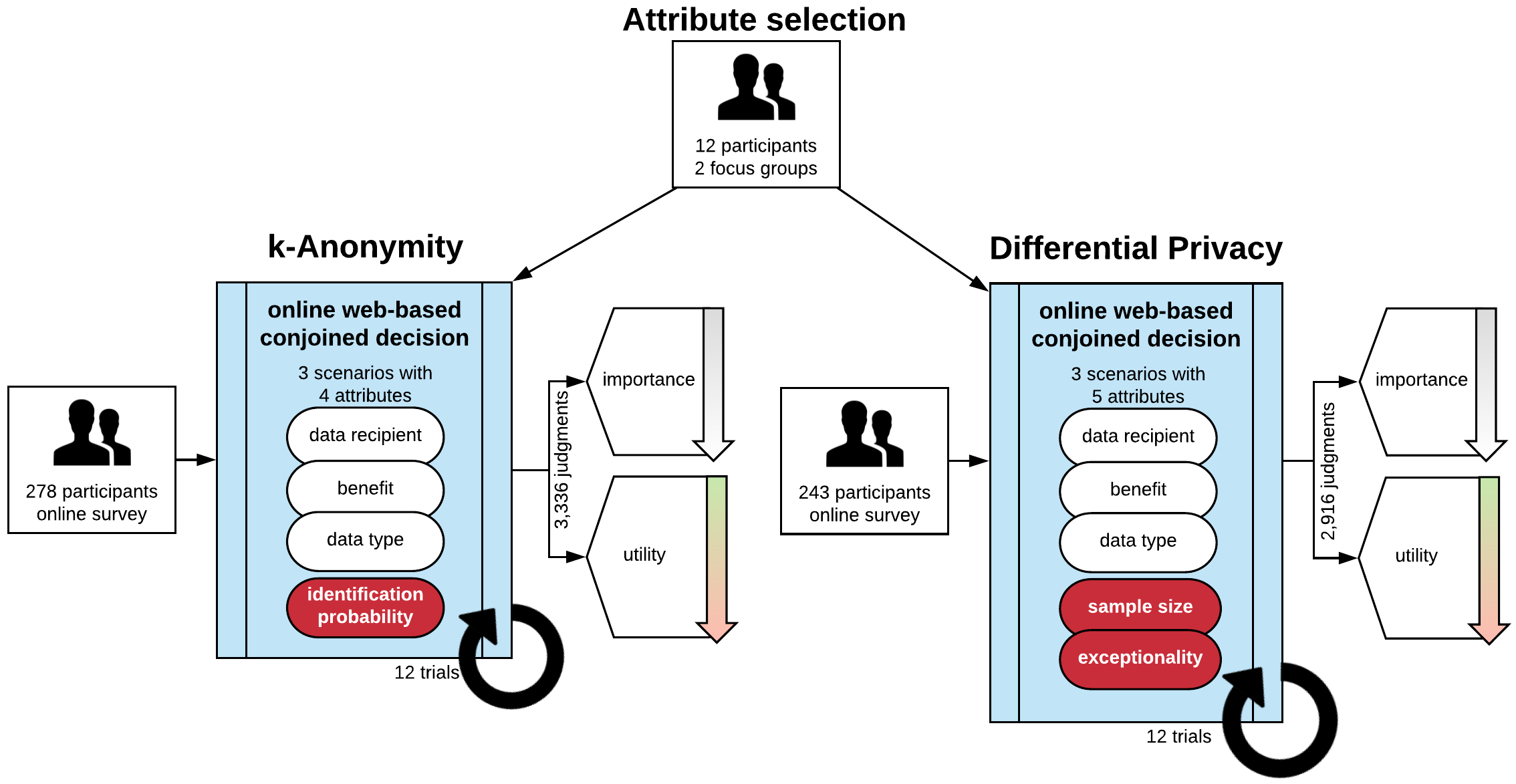}
\caption{Overview of the methodological approach showing both the qualitative focus groups to determine relevant attributes and the two conjoint decision tasks. Attributes in red differ between both studies.}
\label{fig:method}
\end{figure}

In order to understand people's attitudes and opinions about the topic we undertook a three-step empirical approach (see Fig. \ref{fig:method}), combining qualitative (focus groups) and two quantitative procedures (conjoint analyses). The whole study was carried out in Germany, therefore the findings are based on a German perspective on the willingness to share medical data to the public.\footnote{It should be noted that a German perspective on privacy is a distinctive and idiosyncratic one \citep{whitman2004two}, given its history and its law regarding privacy.}

\subsection{Gathering People's Attitudes: The Focus Group Approach}
In a first step, three focus groups were conducted (mixed groups, 18 participants in total, six each, age range from 20--50 years, 50\% female). \added{Participants were recruited among acquaintances, citizens of Aachen, and university students. They reacted to posts distributed in social media and at public campus boards, in which voluntary participants were searched for the topic ``privacy in health.'' The motivation to join the study was high as all participants reported to have a strong interest in taking an active part in the development of a public understanding of privacy issues in the context of digital health services. Focus groups were held in May 2016 and lasted around 2.5 hours. They were carried out on the university campus. Three students who had received prior training moderated each focus group. Before the interviews started, we informed interviewees that participation was voluntary and not gratified. Also, detailed information was given about the purpose and the aim of the study. We also stressed that participants should feel free to comment on the topic and to openly share opinions. Participants were also informed that none of their answers could be referred back to them as persons.}

\added{The focus groups were structured in three parts. First, we introduced the topic ``privacy issues in the context of health and medical systems'' and explained possible consequences in both, societal and individual benefits as well as drawbacks.  In a second part,} 
we asked participants to elaborate their thoughts on the topic, discussing both the benefits and the drawbacks of sharing medical data on the Internet. 
It was impressive to see how fragmented acceptance patterns were in this respect. On the one hand, all participants agreed on the importance of the availability of public health data for the general benefit of societies in terms of health education and medical treatments. On the other hand, participants claimed individuality and intimacy as personal rights and had fervid discussions if and which conditions could be given that would convince them to share their health data for the use in a recommender system. 

\added{In a last step, we asked all participants to note their individual concerns on sticky-notes and then, similar to a card-sorting task, arrange them in order of importance. Then the whole group was asked to discuss and agree on a common order of concerns. Interestingly, all three focus groups agreed regarding the four most important criteria for privacy in digital health services.}

\added{Additional categories that arose only in a single focus group, were the concept of ``dignity in data usage,'' ``dignity of humans is not compatible with any data,'' ``severity of disease'' (i.e. terminal, non-terminal disease), and ``balancing utility and privacy.''}

\added{A small proportion of participants (n=3) claimed, they would not share medical data in any case. Among these three participants one was a computer scientist, and two medium aged users (45 and 47). In contrast, one participant said, she would release all her data, independently of use. ``There is no privacy anyways''. The majority of focus group participants though agreed that there might be acceptable compromises, which depend on the individual preferences for a given context (i.e., illness, data usage, identifiability).
}

On the basis of the focus group findings, the four most important factors were extracted which prominently impact---according to participants---the decision to share medical data on the Internet. First, the type of data is relevant (general health data, physical illnesses, chronic diseases, mental illnesses), also the probability of being identifiable has a major impact as do the benefits offered for data sharing (personal, financial, and general benefit). Finally, the data receiver is extremely important to participants, meaning those persons or entities that receive or use the data (science, health insurances or companies for commercial use of the data). The four attributes and their variations were then transformed into an experimental design of decisions scenarios for the subsequent second study, a quantitative conjoint study which is explained in the next section. 

\subsection{Understanding Decisions to Share Data: The Conjoint Analysis-Approach}
Whenever we want to understand under which circumstances people agree to share their medical data, it is not sufficient to examine the relevant factors in isolation, independently from each other. In reality, such decisions are reached within a given scenario, in which different levels of the factors at hand are prevailing. For example, users could decide to share their medical data, knowing that they are used for the benefit of others suffering from chronic diseases, even though there might be a certain probability of being identified. Likewise, users could refuse to share medical data in case of a company utilizing the data, whom they distrust. From a psychological point of view, such decisions then represent a product of a weighing process. In such cases, the traditional questionnaire approach in which single factors are examined is not suitable, as the findings tell us only to what extent the single levels of the relevant factors are accepted but not the trade-offs in between. 

Therefore, we used another empirical method that allows to identify such trade-offs: The choice-based conjoint analysis approach mimics the complex decision processes in real world scenarios in which users have to evaluate more than one attribute that influences the final decision \citep{luce1964simultaneous}. In the context \replaced{at hand}{here}, the trade-off between sharing medical data for a recommender system vs. keeping one's own privacy was experimentally studied. Methodologically, the presented decision scenarios and trade-offs consist of multiple attributes and differ from each other in the attribute levels. As a result, the \textit{relative importance} of attributes deliver information about which attribute influences the respondents' choice to what degree. Part-worth utilities reflect which attribute level is valued the highest and how much so. For the experimental design, we used a $4\times 4$ respective $5\times 4$ factors matrix (see Table~\ref{tab:tab1}). Levels were chosen from the focus group results to find levels that are sufficiently different from a users' perspective. This approach has been \replaced{successfully}{previously} applied to privacy perceptions in other contexts\deleted{ successfully}~\citep{ziefle2016users}. The findings from such studies help understand how users evaluate criteria in a conjoint-setting. All attributes must be evaluated at the same time. This allows to measure trade-offs more precisely than in disjoint, consecutive measurements (e.g.~anchored rating scales, etc.).
This procedure \replaced{enables}{allows} to derive rules of how much one attribute is worth in terms of the other. Conjoint-analyses \replaced{facilitate understanding}{allow to understand} the middle-ground of attribute levels, and self-correct for reporting-biases, thus they allow to judge decisions more adequately.
In our case, we use the Software Sawtooth Lighthouse Studio 9.5.3 for our survey generation and analyses.


\begin{table}[htb]
\centering
\caption{Five attributes and respective levels of our two study. Study one uses attributes 1--4, study two replaces attribute 4 with 4a and includes attribute 5.}
\label{tab:tab1}
\begin{tabular}{@{}llcccc@{}}
\toprule

\textbf{Attribute}                                                            &  & \multicolumn{4}{c}{\textbf{Levels}}                                                                                                                                                                                       \\ \midrule
\textbf{1. Type of data}                                                         &  & \begin{tabular}[c]{@{}c@{}}General\\ health data\end{tabular} & \begin{tabular}[c]{@{}c@{}}Physical\\ illness\end{tabular} & \begin{tabular}[c]{@{}c@{}}Chronic\\ illness\end{tabular} & \begin{tabular}[c]{@{}c@{}}Mental\\ illness\end{tabular} \\
\textbf{\begin{tabular}[c]{@{}l@{}}2. Benefits of\\ ~ ~ sharing data\end{tabular}}   &  & personal                                                      & financial                                                  & general                                                   &                                                          \\
\textbf{3.~Data reciever}                                                        &  & science                                                       & \begin{tabular}[c]{@{}c@{}}health\\ insurer\end{tabular}   & \begin{tabular}[c]{@{}c@{}}commercial\\ use\end{tabular}  &                                                          \\ 
\textbf{\begin{tabular}[c]{@{}l@{}}4. Identification\\ ~ ~ probability\end{tabular}} &  & 100\%                                                         & 50\%                                                       & 25\%                                                      & 10\%                                                     \\
\textbf{\begin{tabular}[c]{@{}l@{}}4a. Sample size\\ ~ ~ ~probability\end{tabular}} &  & 10                                                         & 100                                                       & 1,000                                                      & 10,0000                                                     \\
\textbf{\begin{tabular}[c]{@{}l@{}}5. Exceptionality\\ ~ ~ probability\end{tabular}} &  & unique\%                                                         & 5\% similar                                                       & 20\% similar                                                    & average                                                      \\
\bottomrule
\end{tabular}
\end{table}

\subsection{Experimental Designs and Decision Scenarios}
Decision scenarios were provided using an online questionnaire. We aimed \replaced{at representing}{to represent} two underlying differences in privacy: 
For k-anonymity, we suggest a risk of identification, and for differential privacy we suggested a sample size and a level of exceptionality. While technically both anonymization procedures are quite different in the underlying idea, it is not yet clear from a social science perspective \deleted{in} how both anonymization procedures are perceived differently and lead to different decisions. 
Therefore, we varied the instructions in line with both anonymity procedures in two consecutive polls using conjoint analyses. The first poll was directed \replaced{at}{to} k-anonymity, the second poll \replaced{at}{to} differential privacy. In both parts, we collected about 250 participants in a wide age range (18--78 years).
Before showing the different instructions in detail, we first report on the common parts of the questionnaire procedure.

\begin{figure}[htb]
\centering
\includegraphics[width=0.7\textwidth]{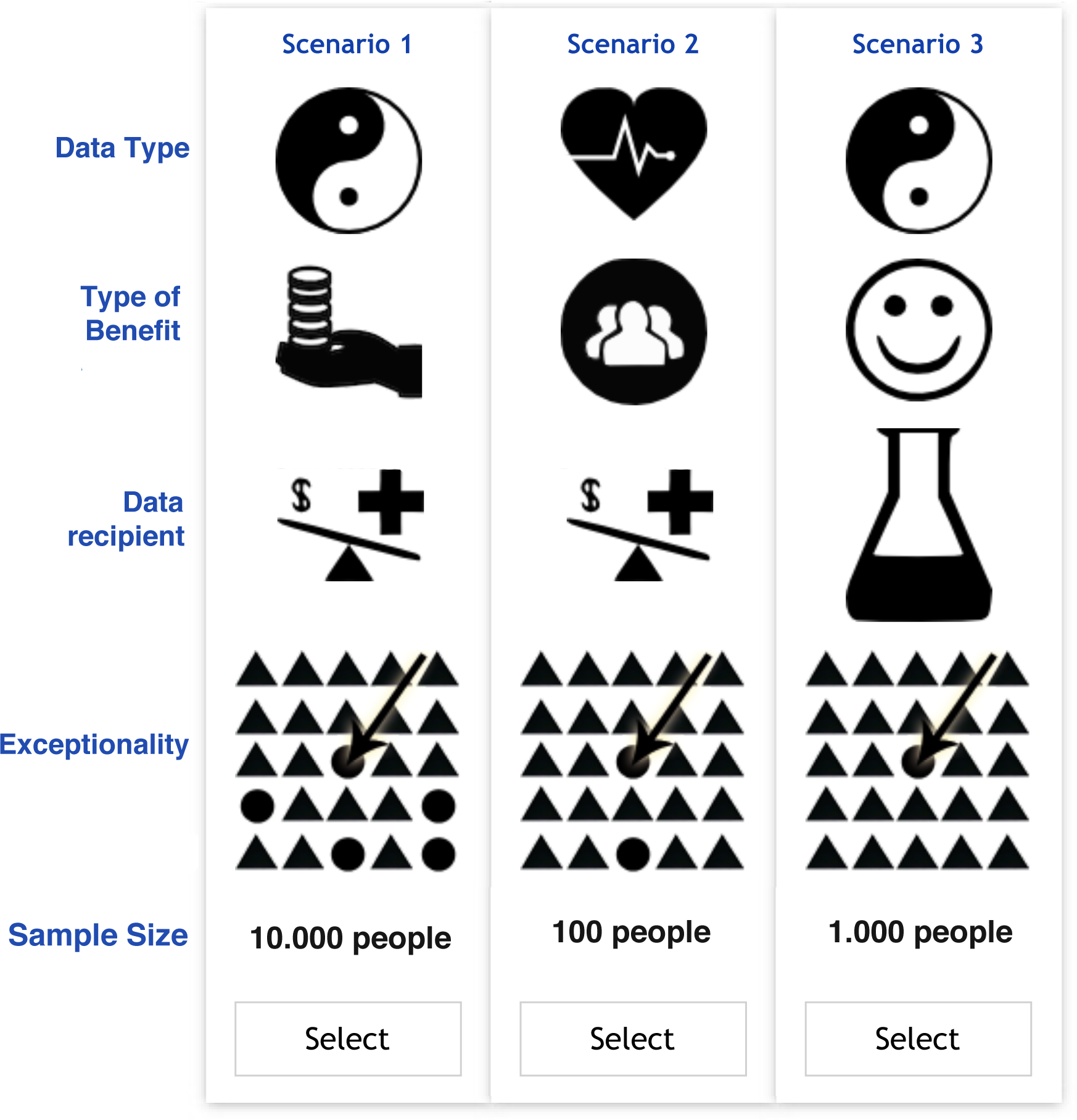}
\caption{Exemplary decision scenario from study 2. Each picture refers to one level of an attribute. Scenario 1 in this case refers to sharing data of a mental illness when the patient is among 20\% similar users in a sample of 10,000 users. The data used in a recommender systems would provide financial benefit for the patient (e.g.~cheaper therapy) and would be stored by a health insurance company. Participants are asked to select their preferred scenario. Actual decision tasks were supported with textual descriptions of attribute levels as tool-tips.}
\label{fig:fig1}
\end{figure}

The questionnaire \deleted{in} was composed using the SSI Web Software by Sawtooth\footnote{The software can be found here: \url{https://www.sawtoothsoftware.com/}} and consisted of three major parts. 
First, participants were introduced into the topic and the reason for the questionnaire (i.e., usage of health data in a recommendation system).
In a second part, demographic data was assessed such as age, gender, health status, and profession. 
Finally, the attributes and their levels were carefully described and instructed, followed by the decision scenarios which were formed out of different levels of the attributes described. In Figure \ref{fig:fig1}, an exemplary scenario choice is illustrated.

A combination of all corresponding levels would have led to 576 ($4\times 4\times 3\times 3\times 4$) possible combinations. As decision tasks are quite demanding, the number of choice tasks was limited to 10 random tasks and 2 fixed tasks. In total, 3,336 conjoint decisions were collected in study one and 2,916 decisions were collected in study 2. A test of design efficiency confirmed that the reduced test design was comparable to the hypothetical orthogonal design. Each choice task consisted of three different combinations of the attributes: type of data, extent of anonymization (two variants), type of benefits, and type of data receiver. 

Participants were instructed to select the scenario they preferred the most. To improve comprehensibility, attribute levels were presented by pictographs; by hovering the mouse over them a tool-tip would provide a textual description. 
To ensure data validity, we removed the top 5\% speeders to \replaced{eliminate}{remove} possible click-through participants. No further validation was applied.

\section{Study 1: k-Anonymity}
In the following, we detail the methodological approach including the description how participants were instructed, the sample, and the results.

\subsection{Participant Instructions}
In order to address k-anonymity, the participants were instructed as follows:

\textit{Patients have a right to decide what will happen with their data. Principally, the society as a whole and every single individual can profit from public health data that are generated on the Internet. What is important is an approach that satisfies the interests of all parties concerned. Here, privacy preserving technologies can help as they anonymize data and thereby detach information from the person. This procedure also reduces the utility of the data as one cannot, for example, link a gender to a specific person anymore. Thus, complete anonymization might not be reasonable in every case. The study aim is to find a solution that adheres to the interests of the data owners (i.e., you as patient) and the ones utilizing the data.} 
\textit{In this questionnaire, we ask for your personal evaluation of different scenarios. Please envision that you have the possibility to share your medical data and also receive the advantages from sharing.}

\subsection{Sample}
Data of 281 participants was analyzed. The sample consisted of 46\% male respondents ($N = 130$) and 54\% female respondents ($N = 151$). The age range was wide, with participants from 18--76 years of age ($M = 39.7$ years, $SD=14.3$). The acquisition of participants for the study occurred through an independent marketing research company in order to reach a wide age distribution, gender equality, and a nation-wide collection of data.
Educational levels varied across participants. When asked about their highest degree of education, the participants \replaced{answered as follows}{represented different educational levels}: 27\% of them reported a university degree ($N = 75$), 28\% ($N = 78$) completed high school, 18\% ($N = 50$) had a vocational education, and 21\% indicated elementary and secondary school graduation. When asked about their current profession, 13\% ($N= 37$) reported to have a medical profession and 87\% ($N = 244$) reported \replaced{other}{to have various} occupations \deleted{fields}, e.g., engineers, administrative assistants, teachers, business economists, translators. Also, participants were asked to indicate their health status (differentiating between chronically ill vs. healthy participants). Regarding health status, 70\% ($N = 196$) reported to be of good health (age range, 18--76 years, $M = 37.9$; $SD= 14.1$), while 30\% ($N = 85$) reported to suffer from a chronic disease (age range 18--65 years, $M = 43.9$; $SD = 13.9$).

\subsection{Results}

The data analysis (estimation of importances and part-worth utilities) was done with the Sawtooth Software (SSI Web, HB, SMRT). In order to identify the main impact factors on users' decision to share their medical data, we calculated the relative importance of each attribute. Then, part-worth utilities were analyzed (on the basis of a hierarchical Bayes multinomial logit model) to understand which attribute was evaluated as most relevant across all decisions and in relation to \added{the} other attributes.

\subsubsection{Relative Importance of Attributes}

\begin{figure}[htb]
\centering
\includegraphics[width=\textwidth]{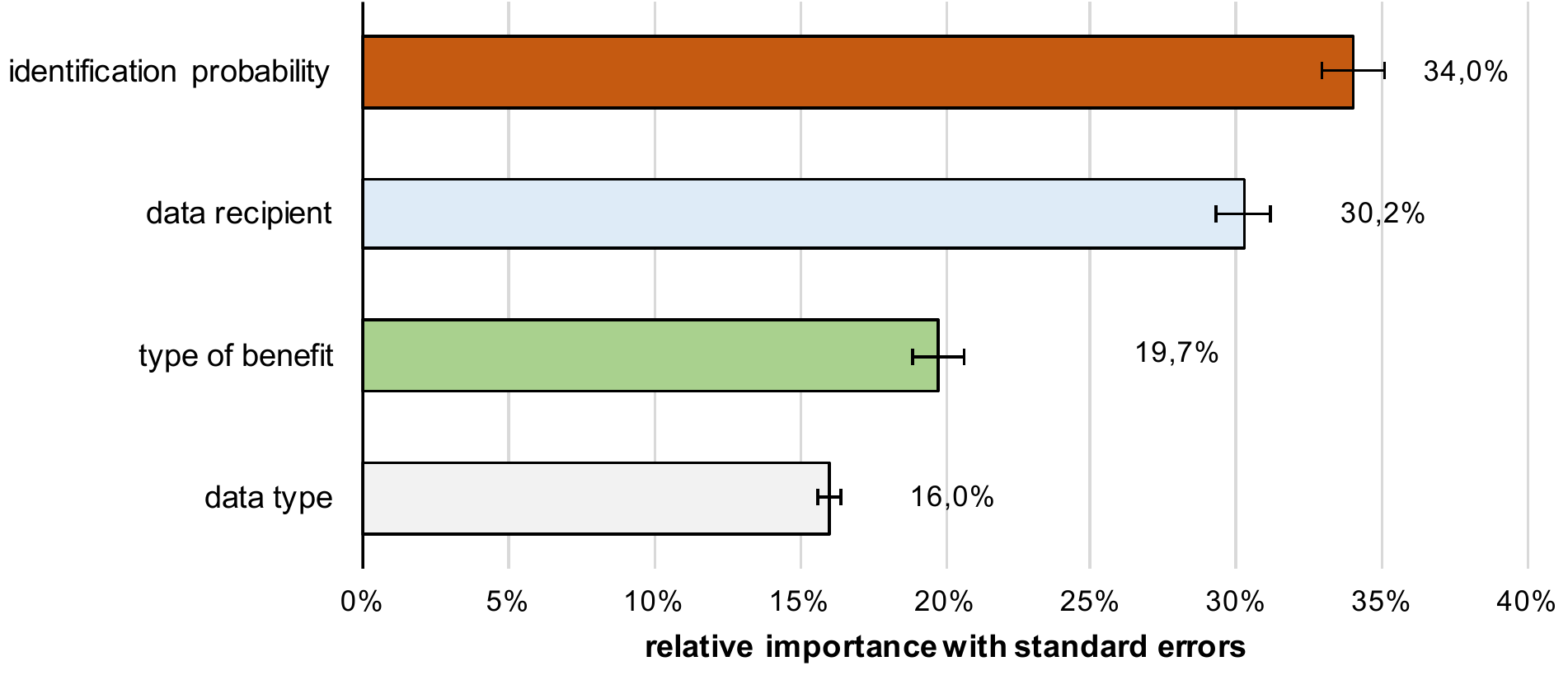}
\caption{Relative importance of attributes for preference. The sum of importances adds to approx.~100\%. Error bars denote standard errors.}
\label{fig:fig2}
\end{figure}

In Figure \ref{fig:fig2}, the relative importance of attributes is pictured. As can be seen, the most important attribute for the decision to share data is identification probability (34\%), followed by the data receiver with a share of (40.8\%); \replaced{thus}{meaning} the entity which is utilizing the data. The type of benefit (19.78\%) and the data type (16\%) are, in comparison with the two most relevant factors, less important \replaced{to}{for} the users' decision to share their data.

In short, we can conclude that it is essential for participants to be sure that the identification probability is low when sharing their medical data. In addition, participants want to know who actually utilizes the data. Interestingly, the data type is evaluated as least important, compared to the other attributes under study. 
The next step is now to analyze the single levels of the attributes. This will be \replaced{presented}{done} in the next section.

\subsubsection{Part-worth Utilities: The Value of Attribute Levels}
\begin{figure}[htb]
\centering
\includegraphics[width=\textwidth]{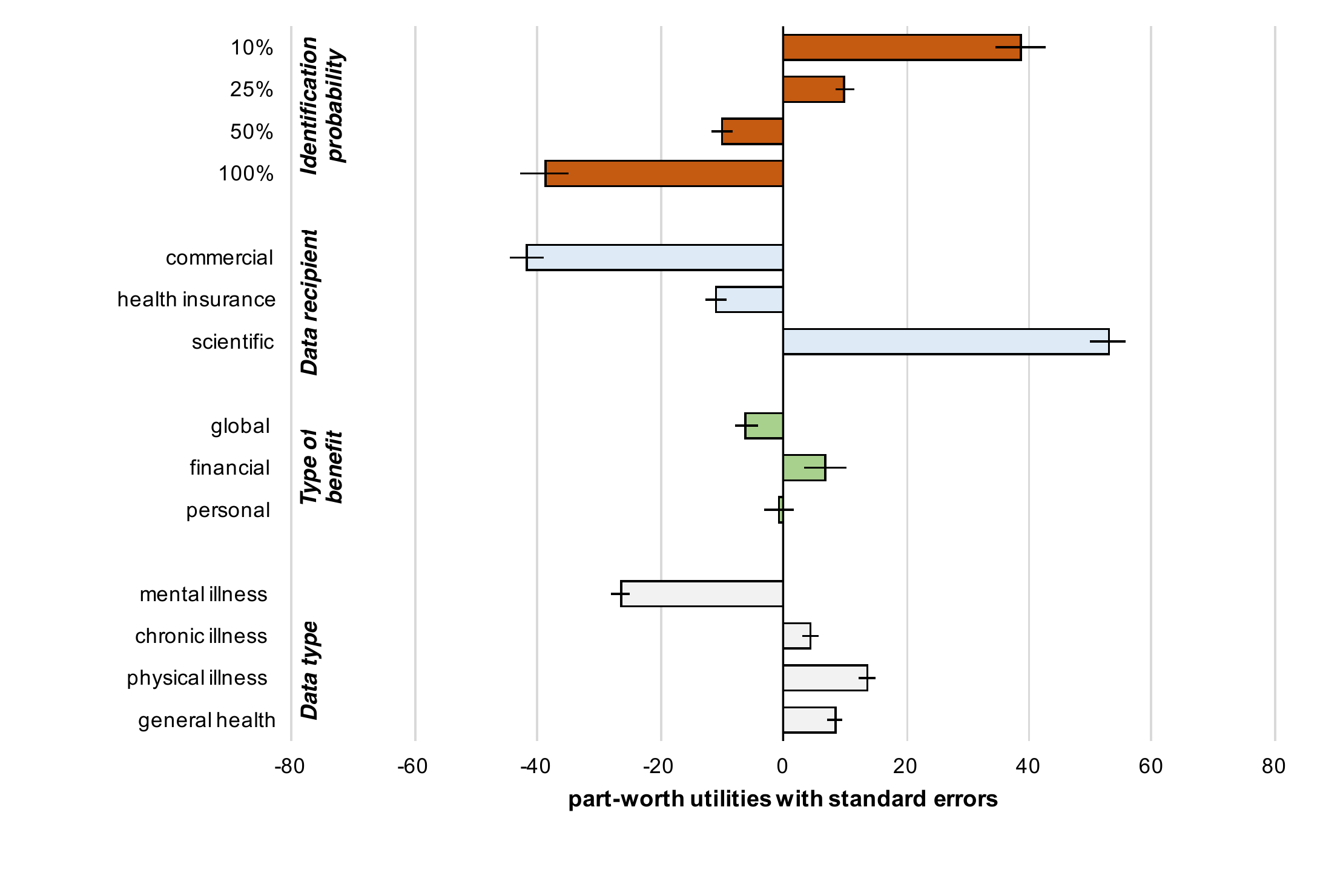}
\caption{Part-worth utilities across attributes and levels. Part-worth utilities add up to zero for each attribute and can not be compared across attributes. Error bars denote standard errors.}
\label{fig:fig3}
\end{figure}

Data are depicted as zero-centered scores in order to show positive and negative preferences across attribute levels. Also, it is possible to identify the best and worst case scenarios across attributes and for both health status groups. In Figure \ref{fig:fig3}, part-worth utilities are presented. 

From Figure \ref{fig:fig3} it becomes evident that the attribute identification probability was the most important attribute, even though its levels were evaluated quite differently. Here, two clear-cut preferences were revealed. On the one hand, the 10\% probability of being identified is quite accepted (38.8). On the other hand, the 100\% identification probability was clearly declined ($-38.7$). Between those poles, the 25\% identification probability was seen still slightly positive ($10$), but \deleted{also} the 50\% identification probability is \added{also} declined ($-10$). 
With respect to the data receiver, all respondents consistently agree to share the data ($31.6$) if \replaced{it is}{data are} used for science, for the increase of knowledge and therapy. Diametrically opposed to that, the participants \replaced{refuse}{disagree} to share their medical data for commercial use ($-20.3$). Apparently, a high distrust in commercial entities is \replaced{prevalent}{prevailing}. The data usage by health insurances is seen slightly negative ($-11.3$). When it comes to benefits that the participants could gain from sharing their data---be it global, financial, or even personal---opinions are comparably neutral. Still\deleted{ however}, the financial benefit was the only one which was perceived \added{as} slightly positive. 
\replaced{Finally,}{A final note reflects} the findings from the attribute data type\replaced{show}{, in which} two striking \replaced{results}{outcomes}\deleted{ become visible}. Data on mental illnesses are not to be shared. In contrast, data on physical illnesses are seen as less sensitive by participants. When it comes to the question whether participants would be willing to share general health data, opinions are divided over this issue; physical illness data is slightly preferred \added{for sharing} over general health or chronic illness data.

\replaced{At}{On} first sight, respondents clearly decline to share their data when a commercial use of the data is intended. Here, it becomes obvious how large the distrust in commercial authorities is and \replaced{that there are grave concerns about}{the far-reaching doubt with} what could happen with the data. Getting back to the focus groups in which the underlying argumentation patterns could be revealed, people stressed the fact that it is not only the lack of trust in what will happen with the personal data, it is also the assumption that data is sold without involving the owners of the data. Against this background, it is quite revealing that participants prefer financial benefits from sharing the data. 

Furthermore, the anonymization extent is a highly critical factor. The probability of being 100\% identifiable is no option for respondents. Finally, data on mental illness needs to be protected and kept in privacy for all respondents. Apparently, the concern that the public finds out about one's mental illness is still much more sensitive in comparison to general health data or data on physical illnesses.

\section{Study 2: Differential Privacy}
In the following \added{sections}, we detail the methodological approach including the description how participants were instructed, the sample, and the results \added{of the second study}.

\subsection{Participant Instructions}
\added{Participants were given the same instructions as the participants from the k-anonymity experiment and additionally} instructed as follows:
\textit{You can decide under which conditions you would like to share your data. Data is shared only in summaries (e.g., averages) for a given purpose. This means that your data could be protected in a data set, as your individual data might not ``stand out'' in the averages that are reported.}

\textit{An example might illustrate this: Assume one wants to find out how much money everyone makes in your neighborhood on average. Your data does not stand out if your salary is near the average. You might also not stand out, when your neighborhood is sufficiently large. In both cases, it cannot easily be determined whether you are a part of the reported mean. It can be said that anonymity is a question of how exceptional you are amongst all other people that are questioned. The more inconspicuous you are the more protected you are.}

\subsection{Sample}
In the second study, the sample consisted of 243 participants. About one half, 49\% ($N = 120$) of the respondents were male and the other 51\% were female ($N = 123$). Participants' age ranged from 18 to 65 years of age, with a mean age of $M = 49.6$ years ($SD = 12.3$). Concerning the educational level, 21 \% ($N = 51$) reported a university degree. 34\% ($N = 82$) had a vocational education, 9\% ($N = 22$) completed high school, and 30\% ($N = 74$) had an elementary or secondary school education. When asked about their profession, only 7\% ($N = 16$) reported to work in the medical context. 
The majority of \added{the} 60\% of respondents ($N = 147$) indicated a good health status (age range 23-65, $M = 51.6$, $SD = 11.7$) whereas 40\% ($N = 96$) reported to \deleted{be to}suffer from a chronic disease (age range 18-65, $M = 48.3$, $SD = 12.6$).

\subsection{Results}

\begin{figure}[htb]
\centering
\includegraphics[width=\textwidth]{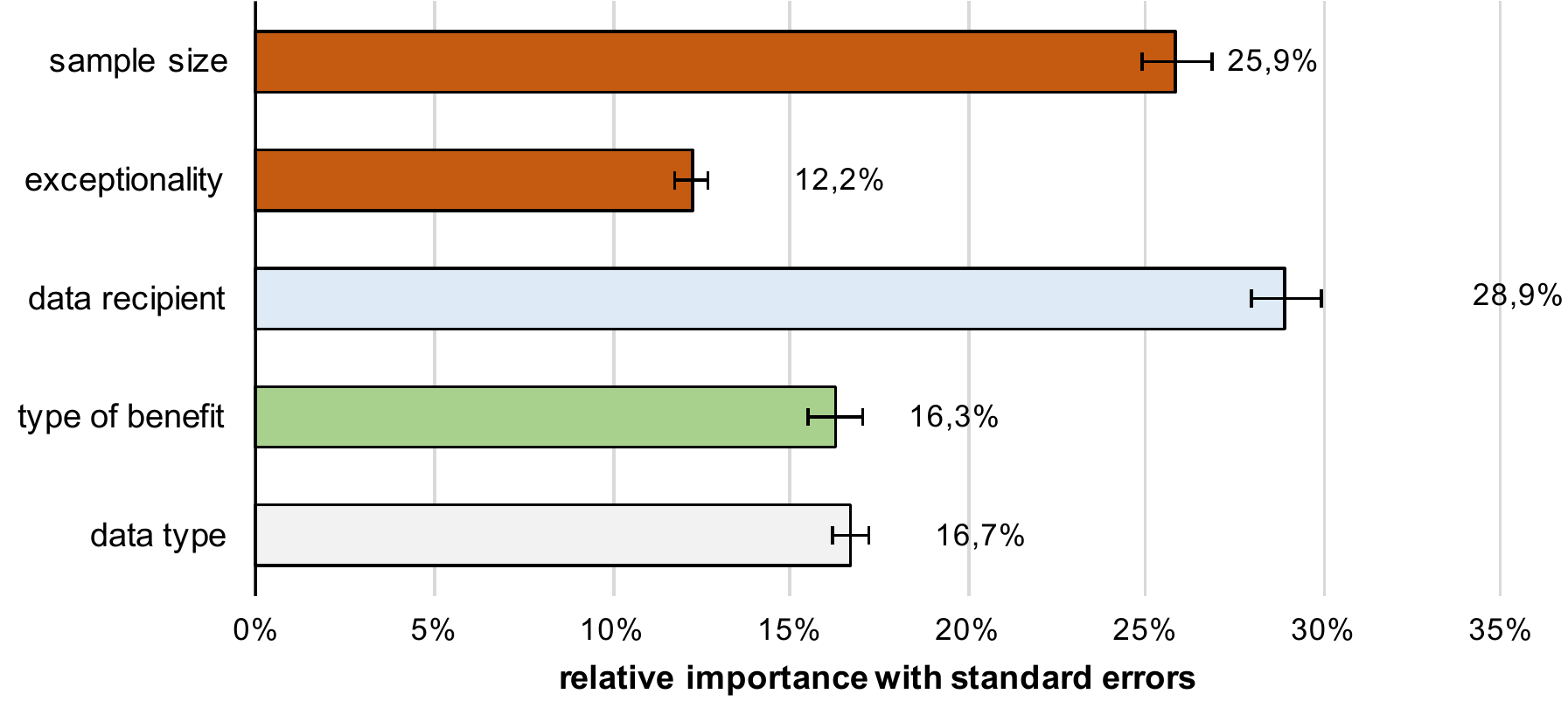}
\caption{Relative importance of attributes for preference. The sum of importances adds to approx.~100\%. Error bars denote standard errors.}
\label{fig:fig2d}
\end{figure}

As in the first experiment, estimation of importances and part-worth utilities was accomplished with the Sawtooth Software (SSI Web, HB, SMRT). In a first step, relative importances of each of the attributes were calculated, reflecting users' decisions to share their medical data (see Fig.~\ref{fig:fig2d}). In a second step, part-worth utilities were analyzed, \deleted{which}showing the weight of each attribute and its levels across all decisions and in relation to other attributes. In Figure~\ref{fig:fig2d}, relative importances are depicted. As in this conjoint study the underlying mechanisms of \deleted{for}differential privacy preserving technologies were instructed, privacy protection was represented by two criteria: one is the sample size and the other is exceptionality (both marked in red in Fig.~\ref{fig:fig2d}). 
The most relevant decision criterion for participants to share their medical data is the data recipient with a share of 28.9\%. Again, it is of high relevance for participants who utilizes their medical data. The second most important criterion is the sample size; thus participants feel protected when their identity is veiled in a sufficiently large crowd. Exceptionality, the degree to which participants are standing out from the average, does have, in contrast, a much lower decision relevance. The type of benefit and the data type appear to have a similar importance (both level at about 16\%).

\begin{figure}[htb]
\centering
\includegraphics[width=\textwidth]{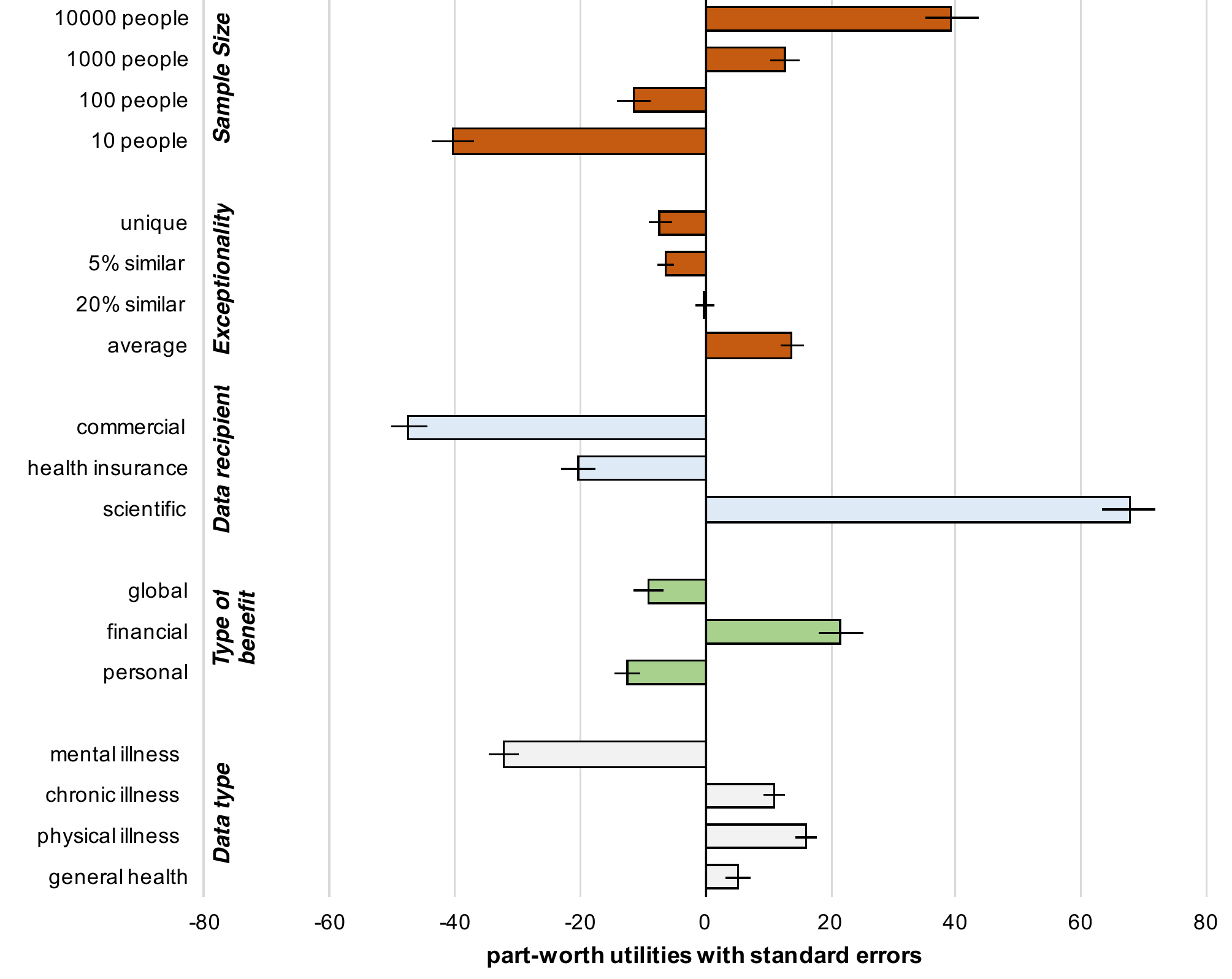}
\caption{Part-worth utilities across attributes and levels. Part-worth utilities add up to zero for each attribute. Error bars denote standard errors.}
\label{fig:fig3d}
\end{figure}

Now, a closer look is directed \replaced{at}{to} part-worth utilities and the respective levels of the single attributes (see Fig.~\ref{fig:fig3d}). To start with the most important attribute, the data receiver, a clear-cut outcome was found. The one and only reason which participants accept as worth for sharing their medical data is science and \replaced{the additional asset}{the surplus} of contributing to public knowledge gain ($67.7$). Health insurances as data receiver are refused ($-20.3$) as \replaced{is}{well as} commercial use with the most negative share ($-47.4$). Standing out among a crowd of 10,000 people is positively evaluated ($39.4$), thus participants would be willing to share their medical data in this case, obviously trusting that they cannot be identified within this sample size. While the next level, standing out among 1,000 people, is also evaluated at least slightly positive\added{ly}, sample sizes of only 100 people or even only 10 people are clearly \replaced{rejected}{declined}, reaching a negative score of -11.5 (100 people) and $-40.4$ in the case of a sample size of 10, respectively. Regarding exceptionality, only the average level was acceptable.  

With respect to the benefits which might be offered for the sharing of medical data, \deleted{the}financial \replaced{gain}{benefit} was perceived positively ($21.6$), while global ($-9$) and personal benefits ($-12.5$) were seen negatively. Thus, global and personal benefits are not powerful enough to motivate participants to share their medical data. As for the data type, findings were, again, clear-cut. Sharing data on mental illnesses is a strong no-go ($-32.2$) while all other data types, physical illnesses ($16.1$), chronic diseases ($10.9$) as well as data on general health conditions ($5.2$), are perceived as less sensitive, \added{all} reaching slightly positive scores. 
From those findings, a best and a worse case scenario can be derived. Participants would be willing to share their medical data whenever their identity is veiled among a sample size of 10,000 people with an average exceptionality (i.e., no exceptionality). Moreover, \replaced{they would share when the}{when} data are used for scientific purposes in combination with financial benefits \deleted{which are} offered for sharing data. Data that would be shared refer to information about (chronic) physical illnesses or general health.
In contrast, the narrative of the worst case is also quite easy to characterize: Absolute no-go conditions for data sharing include the standing out in a very small number of people, \replaced{as well as}{furthermore}, a commercial use of the data, especially when \replaced{those}{data} are related to mental illnesses.

\section{Summary}
In summary, we found a dominant preference for higher anonymity when users are asked about their preferences \replaced{for}{to} shar\added{ing} data \replaced{with}{for} a health recommender system. This is independent from how anonymity is created, either k-anonymity or differential privacy. 
The second most important criterion is the type of usage. Users in both settings prefer \added{a} scientific use \replaced{to a}{, more strongly than} commercial use of data. Lastly, the data type influences preference: Users prefer not to share data on mental illnesses but have little concern to share physical data. The benefit that data sharing has is rather unimportant to users. The strongest preference is for financial benefits. 
Some of the effects might \replaced{originate from} {have their reasons in} our sampling method---a paid market panel in Germany.

\section{Discussion}
In this paper, we empirically analyzed users' perceptions on their willingness to share medical data, the specific conditions under which data sharing is accepted, and also those conditions under which participants choose to stay anonymous. 
To determine the perceived trade-offs between privacy on the one hand and data sharing on the other hand, we used decision scenarios in which different usage conditions were presented. As relevant factors we investigated the type of data, the data receiver, different anonymization conditions (depending on k-anonymity or differential privacy preserving technologies), and benefits that would be given if data \replaced{were}{are} shared.

The findings we present in this article are not surprising. It is expected, when presented with the explicit choice of how privacy should be regarded in a health scenario, that conservative judgments are made. However, in our study we included possible benefits for the participants, or society as a whole. Therefore, sharing health data is not purely seen as something negative. A trade-off decision has to be made. Given these circumstances, the results are important for the design of health recommender systems, as they allow to adjust the goals of future health recommender systems.

Our study is descriptive in nature and does not provide intervention methods, e.g.~how to improve acceptance of data sharing for a given application. Such methods must be carried out with a deeper understanding of the details of the algorithm that is used, e.g., the privacy budget, the concrete benefit, the concrete recipient of the data. Such approaches do not generalize very well, but they can base their decision making on how attitudes shape behavior, as established in this work. For example, if you are a health insurer developing an application for mental health nudging, you must make sure that the negative utility of sharing this type of data with you is compensated for, by either ensuring high privacy constraints or sufficient financial benefits. 

It seems \deleted{to be the case} that the utilization of the immaterial good ``privacy'' for commercial uses is condemned strongly by the participants in the German sample. This is in line with the concept of privacy as a ``personal dignity'' in German culture \citep{whitman2004two}. It is not so much the release of data, as sharing data for scientific purposes is accepted, it is its ill-intended use that causes privacy concerns. 
\replaced{Data does not necessarily have to be anonymized}{It is not that data needs to be anonymized}, \replaced{rather}{ it is more} the usage of data \deleted{that} needs to be controlled (at least from the users' perspective). This attitude is quite ``archaic'' \added{and rooted in human nature} as both healthy and chronically ill people agree on this attribute quite strongly.
This leads to the question  whether this should be achieved by legal or technological means. Possible strategies could include watermarking data~\citep{venugopal2011watermarking} and tracking their use, as well as individualized compensation based on the utility provided by an individual. Such techniques will most certainly be circumvented and add new trade-offs, which is why possibly legal and technological advances are needed.

The individual or global benefit interestingly plays very little into the importance or utility evaluation. One could argue that most participants have little experience with the use of health recommender systems and their utility, so the only option available for judgment is  financial benefit. This option is seen rather negatively, although very little so. No immediate illness is present during this study for which a benefit could be necessary (e.g., a cancer patient looking for a new type of therapy). So this procedure might have caused a general disregard for the different attribute levels. 
In addition, the study was conducted using a market panel, therefore participants themselves provided data for a financial compensation. A different sampling strategy could yield differing results for this attribute.

Anonymity is seen as something positive. The levels in this study were picked quite randomly, and with the intention of being comprehensible without knowledge of the inner workings of a tensor-factorization recommendation algorithm, for example. For this purpose, perceivable levels of being identified in a data set where chosen. These levels could, in principle, directly be applied to k-anonymity procedures. The large spread between the levels exists, however more pronouncedly for healthy people. People with chronic illnesses should have more experience with having to track medical data and sharing it with their doctor. They also have a higher expectation from possible benefits (even though neither of the individual levels for benefits shows a peak). A large amount of health recommender systems is aimed at these patients and this finding is a relief in this regard. Applying recommendation techniques to these types of illnesses does seem to have value to those who need it.

Being identifiable as 1 out of 10 (the most anonymous option in our study) still poses a large de-anonymization risk in a real-world application. Our method just shows how valuable the perception of privacy is in such a setting.  Forcing a user to choose between four options will naturally yield a tendency. Possible extensions of our study should investigate not the immediate risk of being identified but the risks associated with identification. Also, investigations of collateral risk (i.e., risks for relatives and non-releasing subjects) could bear interesting outcomes. 
\added{Interestingly, the perception of privacy is more strongly perceived from sample size than from exceptionality in the differential privacy study. When both criteria are added up, a very similar picture emerges. However, sample size seems to more ``imaginable'' than exceptionality. Here lies one of the problems of differential privacy from a users perspective: How exactly, does being average protect me? Being part of a large crowd seems to more protective for privacy, than being very average in such a crowd. Maybe new methods of communicating the benefits of differential privacy can mitigate this problem.}

The type of data yields one of the most interesting findings for health recommender systems. Users have the least concern about sharing physical illness data (e.g., bone fractures). These illnesses seem to have the lowest perception of being revealing from the users' perspective. This information is helpful in designing privacy-aware health recommender systems as it can be used to configure l-diversity algorithms that anonymize individual data fields or data columns.
Alternatively, testing new recommendation algorithms in this field of health (i.e., physical illnesses) could be promising, as lower rates of rejection are to be expected. Sadly though, mental illnesses are still taboo. And even though digitally mitigated services might help users who are uncomfortable in reaching out in their immediate environment, very little research is conducted in this field of application. In this regard, it should be noted that maybe overcoming this comfort-zone might be one of the key issues in treatment, thus application of recommender systems for mental illnesses must be approached very carefully. The first-do-no-harm principle \citep{ekstrand2016first} must be respected and specific evaluations with patients and recommender systems are \replaced{imperative}{inevitable}.

Overall, privacy plays a different role depending on the usage context, data, and culture. The use of personal health records (PHR) for recommender systems should be applied only when necessary and selectively so. An alternative to a system that uses PHRs could be the use on non-personalized recommendation. When no user-profiles exist, lower de-anonymization risks should result. The utility of such systems should be evaluated for different illnesses and trade-offs and could be generated for specific fields of application.

A different approach could be the use of data mining in encrypted data sets. By applying homomorphic cryptography, algorithms can be designed that have no knowledge of the real data of a patient, but apply, e.g., classification~\citep{bost2015machine} on encrypted data. These procedures work in both training and application.

A large type of benefit typically used in recommender systems was not investigated in our study, namely explanations. Telling a user why a certain therapy is helpful, or why a certain health behavior could help them, might improve confidence in the therapy, thus adherence, and lastly healing. Applying explanations from anonymized data might be less helpful, but further evaluations are required to judge the loss of utility in explanations if they are stated in anonymous fashion. 

Strangely, sharing \deleted{of} fitness related data (walking paths) was not seen as critical in the focus groups, even though this type of data poses high risks of de-anonymization and privacy invasion. The privacy paradox is heavily at play even in health-related settings. Users do share data but value not sharing data. Attitude and behavior are \replaced{discordant}{in disconnect}. This could potentially lead to even stronger feelings of resentment when data is used against the interest of the data sharer. 
Individualized privacy seems to be a good approach for health recommender systems here as well. Such systems should ensure that users understand what they are sharing and how the data is used to ensure \textit{informed} consent. Users should be taught how the sharing of data relates to their privacy \citep{wisniewski2017making} before belated regret can set in. 

This is also one of the dominant questions in privacy research from a social science perspective. Can the user gauge what happens with his data? The approach conducted by computer science researchers is in a way ``typical'' for computer scientists, addressing the problem with another algorithmic solution. However, the underlying problem is not merely a ``data privacy'' or algorithmic problem, it is a ``perceived privacy'' problem---a socio-cognitive problem. Users do follow the privacy paradox, and even computer scientists with experience in privacy research share information about themselves publicly on social media. The question in our work is not necessarily ``what can be done to prevent intrusion'' but rather ``what do informed users want''? Our work assesses what users actually want, not what they should want. Our findings still require intelligent interpretation by an algorithm designer when employed in a real world setting. There is no one-size-fits-all solution as attitudes are additionally shaped by brand names, individual utility expectation, and  Zeitgeist.

Further, the findings from our study have a large cultural bias. German privacy culture is---as previously mentioned---unique and this aspect should be regarded when evaluating our findings for a health recommender system. \cite{li2017cross}, for example, have conducted a cross-cultural evaluation of privacy and found differences in privacy concerns related to the Hofstede cultural dimensions. Individualism in a society leads to lower acceptance of sharing data while collectivism leads to higher rates of acceptance. This type of privacy preference could, on the one hand, help boost health recommender systems by rolling them out in collectivism-centered cultures first. The problem is that such procedures could introduce new bias into the data, reducing their utility, on the other hand. Privacy and utility will remain a trade-off decision.

From a social science point of view, the validation of our data should be performed by a replication study. Especially, as privacy is discussed in the media and politics, this might change public attitudes towards privacy preserving technologies. To accommodate this influence, a longitudinal study (e.g., privacy barometer over time) should address such changes.

\section{Conclusion}
To identify the importance and utility of data sharing and privacy in health recommender systems, we have conducted two conjoint-decision studies with a total of 521 German participants. We have shown that the risk of identification has the strongest influence on utility for a health recommender system presented in such a scenario, followed by the usage context of data. 

This work helps to understand that privacy concerns change depending on what data is collected and how it is used. Our findings can be used to estimate privacy parameters (e.g., $\epsilon$ for differential privacy) for a given scenario. 
For example, a health recommender that suggests nudges for mental health patients requires a much stronger privacy protection if the data will be used by a commercial company than if it will be used in scientific research. 
Further, this study establishes a general method to measure and estimate privacy-utility trade-offs that can be adapted to a specific scenario. It can be employed by the developer of a health recommender system in advance, to estimate the sensitivity of their data, and assess which factors must be addressed to improve acceptance and usage of a system. 
We propose to integrate these findings in future health recommender systems by adjusting privacy-preserving techniques and further development of new strategies to allow individuals to choose their own best-case privacy-utility trade-off. 
\section*{Acknowledgements}
We owe gratitude to Hanna Fleck, Julian Halbey, and Sylvia Kowalewski for their valuable support in the empirical work. Also, we thank Roman Matzutt and Henrik Ziegeldorf from the Chair of Communication and Distributed Systems at RWTH Aachen University for their valuable advice. Further, we would like to thank Chantal Sean Lidynia for proof-reading.
This research has been funded by the Excellence Initiative of German State and Federal Governments (Project NEPTUN, no. OPSF316) and by the German Ministry of Education and Research (Project MyneData, no. KIS1DSD045). The authors thank the German Research Council DFG for the friendly support of the research in the excellence cluster ``Integrative Production Technology in High Wage Countries''.



\section*{References}
\bibliography{mybibfile,savedrecs,privacy}

\pagebreak
\begin{minipage}[c][7cm][t]{0.3\textwidth}
\noindent
\includegraphics[width=0.95\textwidth]{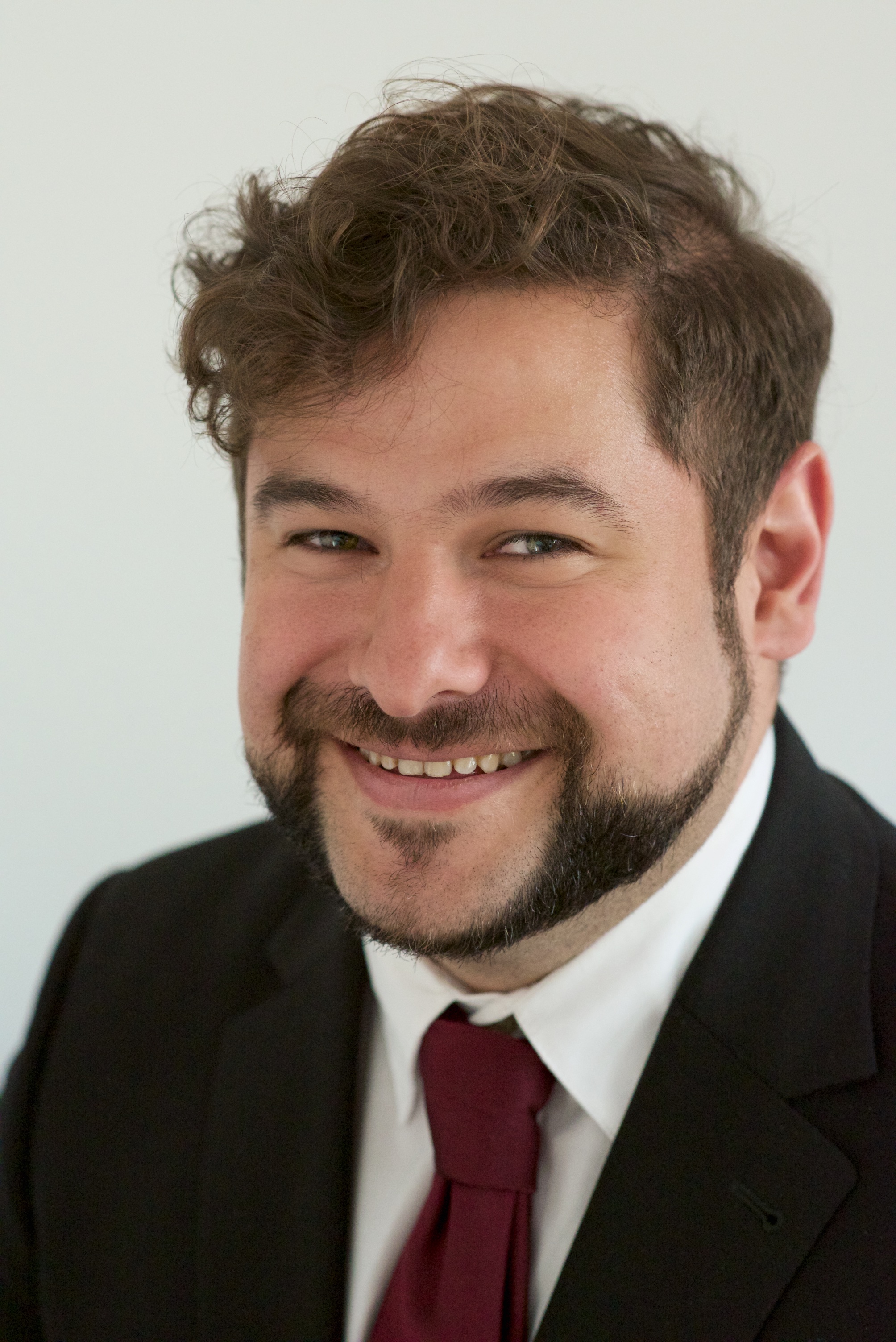}
\end{minipage} \hfill
\begin{minipage}[c][7cm][t]{0.7\textwidth}
Dr. Andr\'{e} Calero Valdez is Senior Researcher at the chair for Communication Science at RWTH Aachen University. He holds a Diplom in Computer Science and a Ph.D. in Psychology and focuses his research on Human-Computer Interaction. His research interest lies in understanding and supporting the process of transforming real-world complex data from novel fields of application (e.g. eHealth, Industrie 4.0) to actionable knowledge. He is active in fields of human factors, information visualization, recommender systems, decision support systems, and machine learning.
\end{minipage}

\vspace{3em}

\begin{minipage}[c][7cm][t]{0.3\textwidth}
\includegraphics[width=0.95\textwidth]{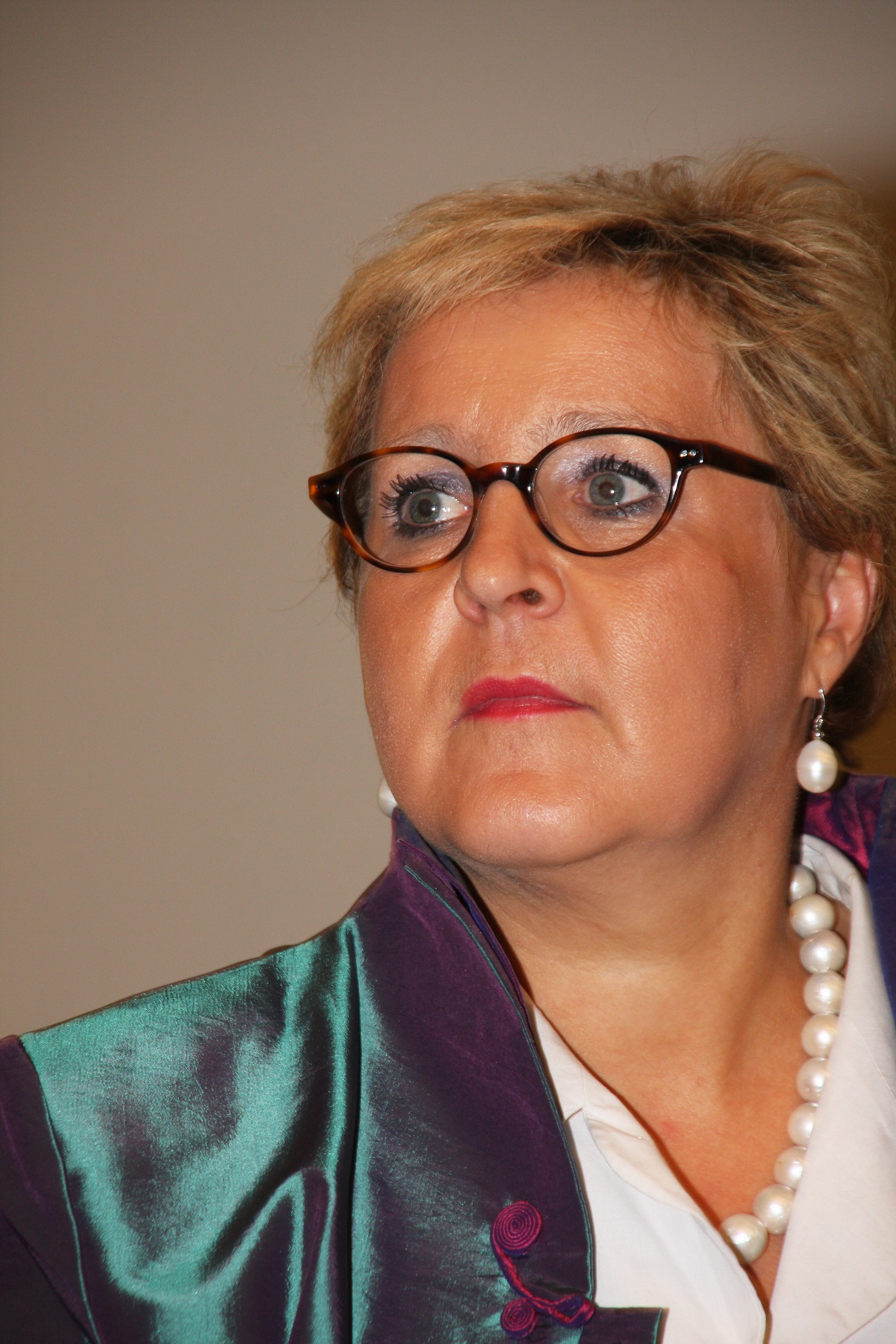}
\end{minipage} \hfill
\begin{minipage}[c][7cm][t]{0.7\textwidth}
\noindent
Prof. Dr. Martina Ziefle is Professor for Communication Science at RWTH Aachen University and director of the Human-Computer Interaction Center at RWTH Aachen University. Her research is directed to human-computer interaction and technology acceptance, taking demands of user diversity into account. She focuses on usability and acceptance of ICT technologies used increasingly in novel contexts (e.g. eHealth). Her main research concern is to shape technology innovation so that technology development is truly balanced with the human factor. In addition to teaching and directing research, Martina Ziefle leads various projects, dealing with interaction and communication of humans with technology.
\end{minipage}

\end{document}